\documentstyle[prb,aps,twocolumn,floats,epsf]{revtex}
\begin{document}
\draft
\twocolumn[\hsize\textwidth\columnwidth\hsize\csname @twocolumnfalse\endcsname
\title{Many-body correlations probed by plasmon-enhanced drag measurements
in double quantum well structures}
\author{H. Noh, S. Zelakiewicz, X. G. Feng, and T. J. Gramila}
\address{Department of Physics, The Pennsylvania State University, 
University Park, PA 16802}
\author{L. N. Pfeiffer and K. W. West}
\address{Bell Labs, Lucent Technologies, Murray Hill, NJ 07974}
\date{\today}
\maketitle
\begin{abstract}
Electron drag measurements of electron-electron scattering rates performed
close to the Fermi temperature are reported.  While evidence of an
enhancement due to plasmons, as was recently predicted
\lbrack K. Flensberg and B. Y.-K. Hu, Phys. Rev. Lett. {\bf 73}, 3572 
(1994)\rbrack, 
is found, important differences with the random-phase
approximation based calculations are
observed. Although static correlation effects likely account for part of
this difference, it is argued that correlation induced multiparticle
excitations must be included to account for the magnitude of the rates and
observed density dependences.
\end{abstract}
]

The plasmons of two-dimensional electron gas (2DEG) systems provide a
valuable platform for the study of electronic many-body correlation
effects.  Fundamental elements of 2DEG plasmons, however, have been
studied\cite{stern,sarma,jain} without resorting to these higher order
effects.  Of particular interest to our work is the system consisting of
two 2DEG's.  In addition to the conventional (optic) mode, with a plasmon
energy whose leading dependence\cite{stern} on wavevector $q$ is $q^{1/2}$,
such a system supports a second plasmon mode\cite{sarma} whose dispersion
relation is linear in $q$ at small $q$.  Recently, theoretical
work\cite{neilson,neilson2,holas} has moved beyond the random phase
approximation(RPA) to consider the role of many body correlation effects.
While the role of these correlations are generally believed to lower the
plasmon frequencies and provide additional damping in the limit of low
densities, a recent calculation\cite{neilson2} has found that their effects
can be important even at relatively high densities.  As both intralayer and
interlayer correlations are possible in the double layer system, 
both  layer spacing and electron density dependencies
are critical to the study of plasmon effects. 

The traditional experimental probe of 2DEG plasmons has been Raman
scattering\cite{pinczuk}.  Such studies have, for example, confirmed the
existence of the acoustic mode in super lattice structures. The electron
densities used in optical studies to date, however, have been high enough
that correlation effects were not evident.  In this paper, we discuss
results of a complementary experimental approach to the study of 2DEG
plasmon modes: electron drag measurements.  The motivation for these
studies results from recent theoretical predictions of Flensberg {\it et al.} 
\cite{flensberg} that plasmons are observable in such measurements, and
preliminary measurements of Eisenstein\cite{eisenstein}.  We find
convincing evidence for the role of plasmons in this transport style
experiment.  This evidence, the overall temperature dependence of the
measurements as well as the general density dependence, confirm recent
measurements observing the enhancement of electron drag by
plasmons\cite{pepper}.  The new results presented here, however, result
from an exploration of a range of densities below that of both the earlier
optical measurements and recent drag measurements, as well as the first
studies of the layer spacing dependence of plasmon enhanced drag.  We find
that for remotely spaced layers a disagreement in magnitude with
theoretical predictions exists, and that there is a greater width in
the peak which develops in the density dependence.  We also find that the
dependence on the relative electron densities for low overall
densities yields a behavior inconsistent with fundamental aspects of RPA
based theoretical predictions; that the maximum enhancement does not occur
at matched densities.  These results are evidence for an
additional plasmon damping mechanism, and we argue that dynamic correlations 
play an important role in the plasmon enhancement of drag.

In electron drag measurements\cite{gramila,solomon,phonon},
electron-electron (e-e) scattering rates are probed through
the momentum transferred, via e-e scattering, from one current carrying
electron system to a second nearby system.  
It has been shown\cite{gramila,solomon} that the ratio of the
voltage induced in one 2DEG to the current in the other, electrically
isolated, 2DEG directly determines the interlayer e-e momentum transfer
rate, $\tau_D^{-1}$.  This determination of the ``drag'' scattering rate is
a consequence of the balance which is established  between
the force of the electric field which develops in the drag layer and
the effective force of interlayer e-e scattering.

A crucial question for this work is how plasmons can play a substantial
role in what is essentially a DC transport measurement, as both plasmon
modes lie above\cite{sarma,jain} the single particle excitation spectrum.
It was recently predicted\cite{flensberg}, however, that plasmons will
have a dramatic effect on drag.  The key element of the
calculations is that coupling of currents to plasmon modes is made possible
by high temperature ($T$) modifications of the single particle excitation
spectrum when $T$ is of order the Fermi temperature, $T_{F}$.  Earlier high
temperature approaches\cite{jauho} neglected these effects and predicted no
plasmon related enhancement.  Because plasmon excitations substantially
alter the 2DEG screening response, drag scattering rates, which result from
screened interlayer interactions, will be substantially
affected.  According to the RPA based calculations\cite{flensberg},
the inter-layer e-e scattering is increased by plasmons for $T >
0.2 T_{F}$, with the increase peaked at $T \sim 0.5 T_{F}$.  The {\it
increase} results from the real part of the effective dielectric constant
going to zero at a plasmon resonance, so screening is {\it reduced}.
Reduction of the effect at higher $T$ is attributed to Landau damping of 
the plasmons.

The samples used in these measurements are mod\-ulation-doped
GaAs/Al$_{0.3}$Ga$_{0.7}$As double quantum well structures grown by
molecular beam epitaxy.  Two samples were used, both with 200\AA \
wide wells, but one with a 225\AA \ barrier between the wells, and the
other 500\AA .  Electron densities for both samples were $1.55 \times
10^{11}\ cm^{-2}$, yielding $T_F \sim$ 65K, with mobilities of all
layers exceeding $2\times10^6 \ cm^2$/Vs.  By applying voltage to an
overall top gate or bias between the layers, the electron densities of
the individual layers could be adjusted, as calibrated by
Shubnikov-de~Haas measurements.  We denote the electron density of the
drag/drive layer as $n_2$/$n_1$.

\begin{figure}[!t]
\begin{center}
\leavevmode
\hbox{%
\epsfysize=3in
\epsffile{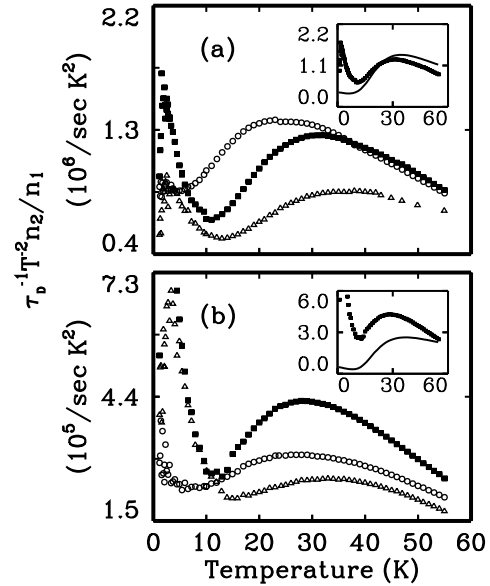}}
\end{center}
\caption{Scaled drag rate $\tau_{D}^{-1}T^{-2}n_{2}/n_{1}$ 
as a function of
temperature for (a) 225 \AA $\ $barrier sample and (b) 500 \AA $\ $barrier
sample. Solid squares are for $n_{2}/n_{1}=1$, circles for $n_{2}/n_{1}=0.5$, 
and triangles for $n_{2}/n_{1}=1.3$. Inset: Theoretical curve based on RPA
(solid line) and the experimental data (solid square) for matched density.}
\label{1}
\end{figure}

Fig.\ref{1} shows the measured temperature dependence of $\tau_{D}^{-1}$
scaled by $T^{2}$ and the density ratio $n_{2}/n_{1}$, for both layer
spacings at various density ratios.  The drag rates show two distinct
maxima, at low and high temperatures.  The peaks  below 10 K
result from a virtual phonon exchange process\cite{phonon}; this work
focuses on the high temperature behavior.  For matched densities
($n_{2}/n_{1}=1$), both samples show an increase in the scaled drag rate
$\tau_{D}^{-1}T^{-2}n_{2}/n_{1}$ near 15K, a maximum near 30K, and a
subsequent decrease with increasing temperature.  This temperature
dependence is expected\cite{flensberg} for plasmon enhancement of
drag, as discussed earlier.
We have also measured the temperature dependences for mismatched densities
(Fig.\ref{1}), and find a shift in the maximum of
$\tau_{D}^{-1}T^{-2}n_{2}/n_{1}$ as the  drag layer density, $n_2$,
is changed.  For $n_2/n_1 = 0.5$, the peak moves to 25K or less in both
samples, and for $n_{2}/n_{1} = 1.3$ it is moves to $\sim$35K.  The
direction and rough magnitude of this shift is consistent with the RPA
plasmon calculations, providing additional evidence for a plasmon based
drag enhancement.

It is important to note, however,  the discrepancies between our
measurements and the published calculations.
Inset in Fig.\ref{1} is a comparison of the measured
$\tau_{D}^{-1}T^{-2}n_{2}/n_{1}$ at matched density with the 
calculations\cite{flensberg} of Flensberg and Hu.  Theoretical rates
appropriate to our sample geometries were derived from 
$T$ dependences for identical 200\AA\ wide quantum wells with densities $n
= 1.5 \times 10^{11}\ cm^{-2}$,
with layer separations 800\AA\  and 400\AA .  A small correction
factor, taken from a plot of numerical results of the explicit spacing
dependence at 40K, was then applied to match to our well center to center
spacings of 700 and 425\AA , respectively.  For the 225\AA\ barrier sample
the measured rates peak at a temperature below that predicted by theory,
with reasonable agreement in magnitude.  For the 500\AA\ barrier sample, 
however, the measured magnitude lies above that
calculated.   The calculations are reasonably
accurate in determining the 
zero temperature limit of Coulomb scattering as determined\cite{phonon} in
these samples (better established, perhaps, for the 225\AA\ sample); a
critical point if magnitude comparisons are to be made.
The data for the 500\AA\ barrier sample show similar behavior to that seen
in earlier measurements\cite{pepper}.

An additional discrepancy with calculations  exists in
the behavior of $\tau_{D}^{-1}T^{-2}n_{2}/n_{1}$ as the densities are
mismatched.    The 500\AA\ barrier sample data indicate a maximum near
matched density, in agreement with the general behavior predicted by
theory for both spacings.  For the sample with more closely spaced
layers, however, we find that $\tau_{D}^{-1}T^{-2}n_{2}/n_{1}$ is
greatest when $n_2$ = 0.5$n_1$.

Two sources for these discrepancies must be considered.  The first is a
shift in the plasmon spectrum induced by correlation effects.  One might
assume that at the  densities of our samples (r$_s \sim 1.4$), such
effects would be small.  Investigations\cite{neilson2} of static many-body
correlations in a double-layer system, however, show that for 250\AA\ layer
spacings, by $r_{s} \geq 5$ the acoustic plasmon is so suppressed 
 that it enters the single particle  continuum and is
destroyed.  While correlation induced changes in the plasmon spectrum for
our samples will be much smaller, they are clearly possible. 
Static correlation effects are found to {\sl suppress}\cite{neilson2} the
energies of the plasmon modes, so the plasmon enhancement should
occur at a lower temperature, providing better agreement with the
observed peak positions in the temperature dependence of 
$\tau_{D}^{-1}T^{-2}n_{2}/n_{1}$ for
both samples.  Small shifts, however, would tend to {\it increase} the
coupling to the single particle spectrum, resulting in a {\sl larger}
drag enhancement. 
Such behavior is found when RPA calculations are modified\cite{pepper}
with the 
Hubbard approximation, which includes intralayer exchange interactions.
The calculations show a shift toward lower peak temperatures and an
increased magnitude, providing better agreement with the measurements of
Ref. 10.
A  net increase in magnitude of the
enhancement, however, is inconsistent with our results for the
225\AA\ barrier sample, where the calculated RPA rates already match the 
observed magnitude. We would anticipate that eventual suppression of  
$\tau_{D}^{-1}T^{-2}n_{2}/n_{1}$ resulting in little net change in magnitude
could also occur, but only for relatively large changes in the plasmon mode
energies, yielding a substantial alteration of the temperature dependences
including the peak position, which is not observed in our measurements.
   
The second possible source for the discrepancies lies in plasmon
dissipation channels only present due to correlation effects.  It is well
known that a conventional RPA approach to plasmons results in fully
undamped modes, precisely because the plasmons lie above the single
particle spectrum.  Realistic estimates of the plasmon widths, therefore,
must include coupling to multiparticle excitations, intrinsically a
correlation effect.  The existence of a multiparticle damping channel could
substantially reduce the degree of drag enhancement provided by plasmons,
potentially offsetting increases due to static correlation effects.
Such damping has been calculated\cite{holas} for plasmons of a single 2DEG
in the density range of this work, but to our knowledge no such
calculations exist for double layer systems.  
Finite temperature effects,
furthermore, would also be expected to play a role in the temperature
range of the current measurements.

\begin{figure}[!t]
\begin{center}
\leavevmode
\hbox{%
\epsfysize=2.5in
\epsffile{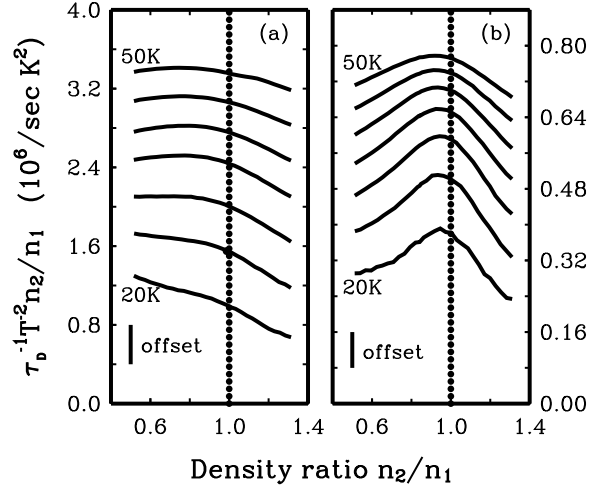}}
\end{center}
\caption{Dependence of the scaled drag rates on the density ratio
$n_{2}/n_{1}$ (drag/drive density), for (a) 225 \AA $\ $barrier sample and
(b) 500 \AA $\ $barrier sample.  Each curve represents the result for every
5 K from 20 K to 50 K and is offset for clarity by the amount shown. A
dotted line indicates  matched density.}
\label{2}
\end{figure}

To further explore these questions, we have measured the detailed
dependence of $\tau_{D}^{-1}T^{-2}n_{2}/n_{1}$ on the relative densities of
the two electron layers at various temperatures.  This measurement probes
the plasmon enhancement while both the plasmons and the single particle
excitations of one layer are changed.  For the data in Fig.\ref{2}, 
only the
drag layer density $n_2$ is varied, while $n_1$ is fixed at 
$1.55 \times 10^{11}\ cm^{-2}$.  
This is, perhaps, a more sensitive probe of the plasmon
enhancement, depending critically\cite{flensberg} on the balance between
exciting the plasmons and damping them through the single particle
excitations.  The response is expected to be sharply peaked at matched
densities.  For the 500\AA\ barrier sample, the scaled drag shows a clear
peak near matched density, which broadens noticeably as $T$ increases.  The
width of this peak, however, is much greater 
than predicted by theory (not shown) at all measured temperatures.
For the 225\AA\ barrier sample, the 
broadening is so large that no clear peak is discernible in the data.

The difference between our measurements and the RPA based calculations is
significant for both samples. The scale for these differences is set by the
observation that the width for the 500\AA\
barrier sample at 30K is very close to that predicted for a 200\AA\
barrier sample.  
Although some suppression of the mode 
should occur due to static correlation effects, this
suppression would, once again, result in an increase in the drag
enhancement.  As any substantial enhancement is inconsistent with our
measurements, we consider it unlikely that correlation induced shifts in
the plasmon spectrum are the sole reason for the broad widths we observe.
Indeed, the recent Hubbard approximation based calculations show no
considerable increase in the width of the density dependence.
Instead, damping of the plasmons via correlation induced multiparticle
excitations must again be considered.  If the plasmon modes are broadened
by multiparticle excitations, then the density dependence would be
broadened, as there would be a distribution, beyond thermal effects, of
energy spacings between the plasmons and the single particle spectrum.  
It is our contention that correlation induced
multiparticle damping plays a significant role in plasmon enhanced drag.

\begin{figure}[!t]
\begin{center}
\leavevmode
\hbox{%
\epsfysize=2.5in
\epsffile{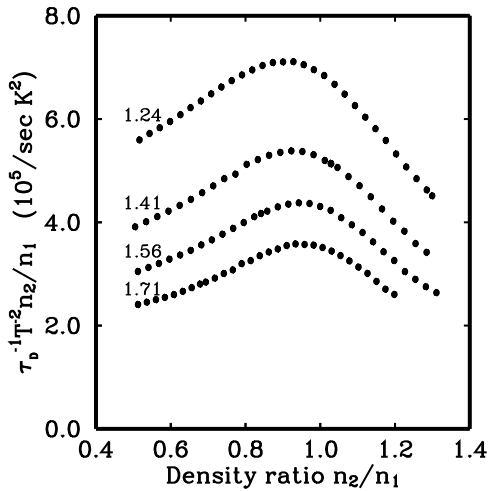}}
\end{center}
\caption{Dependence of the scaled drag rates in 500 \AA $\ $barrier 
sample 
on the density ratio $n_{2}/n_{1}$ for 4 different matched densities at 
30 K. Numbers beside each curve represent the matched 
densities in $10^{11}\ cm^{-2}$.}
\label{3}
\end{figure}

Important evidence for this contention comes from the density dependence of
$\tau_{D}^{-1}T^{-2}n_{2}/n_{1}$, measured for various densities
in the drive layer at 30K.  As seen in Fig.\ref{3}, there is a gradual
increase in the width of the peak with decreasing density.  Most important
in these data, however, is the position of the peak.  
Our measurements show that the maximum in
$\tau_{D}^{-1}T^{-2}n_{2}/n_{1}$ occurs at mismatched densities, and that
the lower the $n_1$, the further the peak is found from matched densities.
Numerical
fits to the data confirm that the deviations are not artifacts of a sloping
background.
For the published
calculations, where the only plasmon damping is coupling to single particle
excitations, the peak $\tau_{D}^{-1}T^{-2}n_{2}/n_{1}$ must
occur\cite{flensberg} at matched densities.  Essential to this result is
the fact that the drag scattering rate depends on the product of the
response of both layers,
combined with a  symmetry resulting
from the fact that the source of the drag enhancement, coupling
between the single particle excitations and plasmons, is also the
damping mechanism. 

The theoretical argument resulting in a peak at matched density fails,
however, if an additional damping mechanism is present.  
As our samples have 20 times higher mobility than a sample of
Ref. 10, it is unlikely that disorder
scattering\cite{das-sarma-guess} provides the additional damping.
Instead,  multiparticle
excitations, which are known to be important in the plasmon damping in
single layer systems\cite{holas} could provide the additional damping.
Correlation effects, furthermore, are widely recognized to become greater
at lower electron densities.  The observed increase in the deviation of the
peak from matched density as $n_1$ is lowered appears, therefore, to be
consistent with a multiparticle excitation damping mechanism.  Although 
the overall effect of plasmon enhanced drag is related
to single particle excitations of plasmon modes, the observation of a peak
in $\tau_{D}^{-1}T^{-2}n_{2}/n_{1}$ at mismatched densities reveals an
additional damping channel.  This is a strong indication that correlation
effects, including the role of multiparticle  excitations in plasmon damping, 
must be included to explain our measurements.

In summary, we have observed an enhancement by plasmons of the interlayer
e-e scattering rates probed by  electron drag measurements.  While our
results agree, in general, with RPA based calculations, we find differences
in the detailed temperature dependences and in the density dependences.
These differences clearly confirm the need to include correlation
effects in calculations of plasmon enhanced drag.  While changes in the
plasmon dispersion curve due to static correlation effects may account for
part of the difference in temperature dependence, we expect such corrections
to cause greater discrepancies in the magnitude of the scattering
rates.   We have proposed that dynamic
correlation effects must also be included to explain our observations.
These multiparticle excitations would provide additional damping of the
plasmons, which  would be consistent with the magnitude of the
scattering rates, the large width of the peak in the rates as a function of
density, and the observation that the maximum in the scaled drag rates
does not occur at matched densities.  A full examination of such effects
awaits further detailed theoretical investigation.

This work was supported by the NSF through grant DMR-9503080, by
the Alfred P. Sloan Foundation,  and by
the Research Corporations Cottrell Scholar program.

\end{document}